\documentclass[fleqn]{revtex4}
\usepackage{amssymb,amsmath}
\usepackage{verbatim}

\begin{document}

\title{Polarizability of the kaonic helium atom}

\author{D.T.~Aznabayev}
\affiliation{Bogolyubov Laboratory of Theoretical Physics, Joint Institute for Nuclear Research, Dubna 141980, Russia}
\affiliation{Al-Farabi Kazakh National University, 050038, Almaty, Kazakhstan}
\affiliation{The Institute of Nuclear Physics, Ministry of Energy of the Republic of Kazakhstan, 050032 Almaty, Kazakhstan}

\author{A.K.~Bekbaev}
\affiliation{The Institute of Nuclear Physics, Ministry of Energy of the Republic of Kazakhstan, 050032 Almaty, Kazakhstan}

\author{Vladimir I.~Korobov}
\affiliation{Bogolyubov Laboratory of Theoretical Physics, Joint Institute for Nuclear Research, Dubna 141980, Russia}



\begin{abstract}
The static dipole polarizability of metastable states in the kaonic helium atoms is studied. We use the complex coordinate rotation method to properly account for the resonant nature of the states. Our calculations show that some of the states are not stable with respect to collisional quenching in a dense helium target and should not be detected in the experiment.
\end{abstract}

\pacs{31.15.A-, 31.30.jf, 31.15.xt}

\maketitle

\section{Introduction}

In 1989 the first direct observation of metastable states of a negative kaon captured by a helium atom had been reported \cite{Yamazaki89}. Interest to this exotic system was revived after the discovery of laser transitions in pionic helium \cite{Nature20}. This makes it possible to improve the determination of the atomic mass of a kaon (pion) using laser spectroscopy \cite{PRA14_first}.

In order to capture kaons into metastable states, the helium target in the experiment must be dense enough to slow down and trap the high-energy kaons, kaonic lifetime at rest is $\tau_{\text{K}^-}=12.4$ ns. In a dense target, the lifetimes of bound states becomes shorter \cite{He-pi_polarizability,Hori15}. In practice, systematic effects such as shift and broadening of the resonance line due to atomic collisions, collisional shortening of the lifetime of states may prevent the experiment from achieving assigned precision. One of the important characteristics of He$^+\text{K}^{-}$ atomic states that can give us some information for precise measurement of specific transitions is the electric dipole polarizability. The case where the state has an anomalously large polarizability and thus can become unstable under experimental conditions is of special importance. Thus the calculated in this work data can be a good guide for the correct choice of transitions for precision laser spectroscopy of kaonic helium atoms.

The aim of our work is to calculate the polarizability properties of metastable states of kaonic $^4$He atoms that are of potential interest. Atomic units are used throughout.

\section{Theory}

\subsection{Wave function}

The kaonic helium atom consist of three interacting particles: K$^-$, electron, and helium nucleus. The strong interaction between K$^-$ and helium nucleus is suppressed due to the centrifugal barrier (the angular momentum of a kaonic orbital $l\approx27$) and may be completely neglected.

The nonrelativistic Hamiltonian of a three-body system is taken in a form
\begin{equation}\label{Hamiltonian}
H =
   -\frac{1}{2\mu_1}\nabla^2_{r_1}
   -\frac{1}{2\mu_2}\nabla^2_{r_2}
   -\frac{1}{M}\nabla_{r_1}\cdot\nabla_{r_2}
   -\frac{Z}{r_1}-\frac{Z}{r_2}+\frac{1}{r_{12}}\,,
\end{equation}
where $\mathbf{r}_1$ and $\mathbf{r}_2$ are position vectors for two negative particles with respect to nucleus, $\mathbf{r}_{12}=\mathbf{r}_2-\mathbf{r}_1$, $\mu_1=Mm_1/(M+m_1)$ and $\mu_2=Mm_2/(M+m_2)$ are reduced masses, $M$ is a mass of helium nucleus, and $Z=2$ is the nuclear charge. We assume that $m_1=m_{\text{K}^-}$ is a mass of a negative kaon and $m_2=m_e=1$. The following values of particle masses were adopted in the numerical calculationsns: $M=m_{\text{He}}=5495.8852754m_e$ and $m_{\text{K}^-}=966.11m_e$.

Kaonic helium presents a quasi-adiabatic system with a heavy kaon orbiting over the helium nucleus with a velocity of about 30 times slower than a velocity of the remaining electron. On the other hand, it may be described as an atomic system with an electron in its ground state: $\psi_{1s}$, while the kaon occupies nearly circular orbital with principal and orbital quantum numbers, $n$ and $l$ \cite{Obreshkov03}. Due to interaction between electron and kaon these quantum numbers are not exact and the wave function is determined by the total angular orbital momentum $L$ and the excitation (or vibrational) quantum number $v$, which are related to the atomic one as follows: $L=l$, $v=n-l-1$.

In our calculations we use a variational expansion based on exponentials with randomly generated parameters \cite{var00}. The wave function is expressed:
\begin{equation}\label{Schwexp}
\begin{array}{@{}l}\displaystyle
\Psi_L(l_1,l_2)=\sum_{k=1}^{\infty}\Bigl\{U_k{\rm{Re}}[e^{-\alpha_kr_1-\beta_kr_2-\gamma_kr_{12}}]
  +W_k{\rm{Im}}[e^{-\alpha_kR-\beta_kr_1-\gamma_kr_2}]\Bigr\}
     \mathcal{Y}^{l_1,l_2}_{LM}(\mathbf{r}_1,\mathbf{r}_2)\,,
\end{array}
\end{equation}
where ${\cal{Y}}^{l_1,l_2}_{LM}(\mathbf{r}_1,\mathbf{r}_2)$ are the solid bipolar harmonics as defined in Ref. \cite{Var88},
\[
\mathcal{Y}^{l_1,l_2}_{LM}(\mathbf{r}_1,\mathbf{r}_2) =
   r_1^{l_1}r_2^{l_2}\left\{Y_{l_1}\otimes Y_{l_2}\right\}_{LM},
\]
and $L$ is the total orbital angular momentum of a state. Complex parameters $\alpha_k$, $\beta_k$, and $\gamma_k$ are generated in a quasirandom manner \cite{var99}:
\begin{equation}
\begin{array}{@{}l}\displaystyle
\alpha_k =
   \left[\left\lfloor\frac{1}{2}k(k+1)\sqrt{p_{\alpha}}\right\rfloor(A_2-A_1)+A_1\right]
\\[3mm]\displaystyle\hspace{10mm}
   +i\left[\left\lfloor\frac{1}{2}k(k+1)\sqrt{q_{\alpha}}\right\rfloor(A'_2-A'_1)+A'_1\right]\,,
\end{array}
\end{equation}
where $\lfloor{x}\rfloor$ designates the fractional part of $x$, $p_{\alpha}$ and $q_{\alpha}$ are some prime numbers, and $[A_1,A_2]$ and $[A'_1,A'_2]$ are real variational intervals, which need to be optimized. Parameters $\beta_k$ and $\gamma_k$ are obtained in a similar way.

We skip discussion of the Complex Coordinate Rotation (CCR) method, the basic concept is reviewed in \cite{Ho83}, for particular details of CCR in application to the considered problem we refer readers to \cite{He-pi_polarizability}.

\subsection{Polarizabilities}

The interaction with an external electric field in the dipole approximation is expressed by
\begin{equation}
V_p = -\boldsymbol{\mathcal{E}}\cdot{\bf d},
\qquad
{\bf d} = e\bigl(Z\,\mathbf{R}_{\text{He}}-\mathbf{R}_{\text{K}^-}-\mathbf{R}_e\bigr),
\end{equation}
where $\mathbf{d}$ is the electric dipole moment of the three particles with respect to the center of mass of the system.

The change of energy due to polarizability of molecular ions is
\begin{equation}
\begin{array}{@{}l}\displaystyle
E_p^{(2)} =
   \langle\Psi_0|V_p(E_0-H_0)^{-1}V_p|\Psi_0\rangle
\\[2mm]\displaystyle\hspace{8mm}
 = E^iE^j \langle\Psi_0|d^i(E_0-H_0)^{-1}d^j|\Psi_0\rangle
 = -\frac{1}{2}\alpha_d^{ij}\mathcal{E}^i\mathcal{E}^j,
\end{array}
\end{equation}
where $\alpha^{ij}$ is the polarizability tensor of rank 2,
\begin{equation}
\alpha_d^{ij}=-2\langle\Psi_0|d^i(E_0-H_0)^{-1}d^j|\Psi_0\rangle.
\end{equation}

Then $\alpha_d^{ij}$ is reduced to scalar, $\alpha_s$, and tensor, $\alpha_t$, terms \cite{Landau}, which can be expressed by the following quantities (we follow notation of \cite{Sch14}) corresponding to the possible values of $L'$ for the angular momentum of the intermediate state: $L'\!=\!L\!\pm\!1$, or $L'\!=\!L$.
\begin{equation}
\begin{array}{@{}l}
\displaystyle a_+= \;\;\frac{2}{2L+1}
\sum_n\frac{\langle 0L\|\mathbf{d}\|n(L\!+\!1) \rangle
            \langle n(L\!+\!1)\|\mathbf{d}\|0L \rangle}{E_0-E_n},\\[4mm]
\displaystyle a_0=-\frac{2}{2L+1}
\sum_n\frac{\langle 0L\|\mathbf{d}\|nL \rangle
            \langle nL\|\mathbf{d}\|0L \rangle}{E_0-E_n},\\[4mm]
\displaystyle a_-= \;\;\frac{2}{2L+1}
\sum_n\frac{\langle 0L\|\mathbf{d}\|n(L\!-\!1) \rangle
            \langle n(L\!-\!1)\|\mathbf{d}\|0L \rangle}{E_0-E_n}.
\end{array}
\end{equation}
where $E_n$ is the energy of the intermediate state $|nL'\rangle$. The polarizability tensor operator on a subspace of fixed total orbital angular momentum $L$ now may be expressed:
\begin{equation}\label{alpha:op}
\alpha_d^{ij} = \alpha_s+\alpha_t
                \left[L^iL^j+L^jL^i-\frac{2}{3}\mathbf{L}^2\right],
\end{equation}
where
\begin{equation}\label{alpha2}
\begin{array}{@{}l}
\displaystyle
\alpha_s^{} = \frac{1}{3}\bigl[a_++a_0+a_-\bigr],
\\[4mm] \displaystyle
\alpha_t^{} = -\frac{a_+}{2(L+1)(2L+3)}
              +\frac{a_0}{2L(L+1)}-\frac{a_-}{2L(2L-1)}\>.
\end{array}
\end{equation}
The numerical calculation of the last two quantities, $\alpha_s$ and $\alpha_t$, is the main goal of the present work. In general, these coefficients are complex numbers, and the imaginary part may be interpreted as the contribution to the Auger width of a particular state of azimuthal quantum number $M$.

\section{Results and Discussion}

\begin{table}[t]
\caption{Nonrelativistic energies $E_{nr}$ (in a.u.), Auger widths $\Gamma$ (in a.u.), scalar $\alpha_s$ and tensor $\alpha_t$ polarizabilities for the $^4\mbox{He}^+K^-$ atom.}\label{tab:He4}
\begin{center}
\begin{tabular}{c@{\quad}l@{\quad}l@{\quad}r@{\quad}r@{\hspace{2mm}}}
\hline\hline
\vrule width 0pt height 11pt 
state & $~~~~~~~~~~E_{nr}$ & $~~~~\Gamma/2$ & $\alpha_s~~~~~$ & $\alpha_t\times10^3~~~~~$ \\
\hline\hline
\vrule width 0pt height 11pt 
$(25,23)$ & $-$3.4024865(2)     & $2.56\cdot10^{-5}$ &  $3.58576 + i\;0.33877$ & $-1.39927 - i\;0.51535$\\[1.5mm]
$(26,24)$ & $-$3.38789866633(1) & $3.11\cdot10^{-9}$ & $-1.59291 + i\;1.69676$ & $0.41197 - i\;2.37835$ \\
$(26,25)$ & $-$3.20622306613(1) & $4.3\cdot10^{-10}$ &  $1.34155 + i\;0.00028$ & $0.43132 - i\;0.00035$ \\[1.5mm]
$(27,25)$ & $-$3.06540046242(1) & $7.97\cdot10^{-9}$ &  $1.02953 + i\;0.00023$ & $0.64852 - i\;0.00026$ \\[1.5mm]
$(28,25)$ & $-$2.946595302(3)   & $4.950\cdot10^{-6}$&  $1.25817 - i\;0.65030$ & $0.48307 + i\;0.62204$ \\
$(28,26)$ & $-$2.931443594442(1)& $6.54\cdot10^{-10}$&  $0.75587 + i\;0.00060$ & $0.72171 - i\;0.00068$ \\[1.5mm]
$(29,25)$ & $-$2.8455617(1)     & $6.98\cdot10^{-5}$ & $-1.77316 + i\;2.66409$ & $25.7724 - i\;8.44617$ \\
$(29,26)$ & $-$2.83199839979(1) & $5.54\cdot10^{-9}$ &  $0.43025 + i\;0.00028$ & $1.07217 - i\;0.00030$ \\[1.5mm]
$(30,26)$ & $-$2.7482558446(1)  & $3.4\cdot10^{-9}$  & $-641.028 + i\;93.375~$ & $584.298 - i\;77.294~$ \\
$(30,27)$ & $-$2.734026229098(2)& $2.34\cdot10^{-10}$&  $0.14811 + i\;0.00028$ & $1.22025 - i\;0.00028$ \\[1.5mm]
$(31,27)$ & $-$2.66443660325(1) & $1.21\cdot10^{-9}$ & $-0.09273 + i\;0.00771$ & $1.66610 - i\;0.00808$ \\
\hline\hline
\end{tabular}
\end{center}
\end{table}

In our calculations, the initial states were obtained using the variational expansion (\ref{Schwexp}) by the Complex Coordinate Rotation method. The basis sets up to $N=3000$ functions had been used to get the complex energy of a state. The relative precision achieved is $10^{-10}$--$10^{-14}$, depending primarily on the Auger width of the state. The intermediate states were obtained using the same variational expansion for states of angular momentum $L'=L,L\pm1$. The number of basis functions for the intermediate states was taken to be of the same size as for an initial state, namely, $N'=3000$ for each $L'$. The states are labeled by two numbers $(n,l)$, the principal quantum number $n$ and the orbital angular momentum $l$ of the kaonic orbital, as explained in Sec.~2.1. Convergence was studied in \cite{He-pi_polarizability} and is not considered here.

The final results are presented in Table \ref{tab:He4}. All the given digits of the polarizability coefficients are significant with one exception: those states that have large imaginary part and expected to be unstable in a dense media are calculated with less precision. It is clearly seen that for the case of "regular" states (with small imaginary part in polarizability) the tensor polarizability grows rapidly with increase of principal quantum number $n$ of kaonic orbital. It means that the geometry of the states become less and less spherically symmetric, what should in its turn increase the Stark quenching of the high $n$ states.

The anomalous behaviour of the states with a large imaginary part of the polarizability can be explained by a strong correlation with excited electron "Rydberg" states that arise in the immediate vicinity of the initial state on the Riemann surface of complex energy. These excited states have much broader Auger width and and this proximity begins to affect the polarizability and stability of the initial atomic state. In the case of antiprotonic helium, such states were discussed in \cite{Yamazaki2002}. In \cite{Yamaguchi04}, it was found that in dense targets these states have much shorter lifetimes than predicted by the calculated Auger lifetime assuming an isolated atom.

Thus, from the obtained numerical results it can be concluded that the $(30,26)$ state should not be observed in the experiment, while the states: $(25,23)$, $(26,24)$, $(28,25)$, and $(29,25)$ might be detectable, but can be very sensitive to the target density and difficult to measure accurately..

\section*{Acknowledgements}

We are indebted to M.~Hori for many fruitful discussions. This research has been funded by the Science Committee of the Ministry of Science and Higher Education of the Republic of Kazakhstan (Grant No. AP19175613).


\end{document}